\documentclass[12pt]{article}

\begin{document}

\begin{titlepage}

\begin{flushright}
IUHET-494\\
\end{flushright}
\vskip 2.5cm

\begin{center}
{\Large \bf Lorentz Violation and the Yukawa Potential}
\end{center}

\vspace{1ex}

\begin{center}
{\large B. Altschul\footnote{{\tt baltschu@indiana.edu}}}

\vspace{5mm}
{\sl Department of Physics} \\
{\sl Indiana University} \\
{\sl Bloomington, IN 47405 USA} \\

\end{center}

\vspace{2.5ex}

\medskip

\centerline {\bf Abstract}

\bigskip

We analyze Lorentz violations in the bosonic sector of a Yukawa-type quantum field
theory. The nonrelativistic potential may be determined to all orders in the
Lorentz violation, and we find that
only specific types of modifications to the normal Yukawa potential
can be generated. The influence of this modified potential on scattering and
bounds states is calculated.
These results could be relevant to the search for new macroscopic forces, which may
not necessarily be Lorentz invariant.

\bigskip

\end{titlepage}

\newpage

There is presently quite a bit of interest in the possibility that Lorentz
symmetry may not be exact in nature. If Lorentz violation does exist, it would be
a very important clue regarding the nature of Planck-scale physics. There are
a number of possible ways that Lorentz violation could arise, including
spontaneous violations of the symmetry in
string theory~\cite{ref-kost18,ref-kost19} and elsewhere~\cite{ref-altschul5},
mechanisms in loop quantum gravity~\cite{ref-gambini,ref-alfaro} and
non-commutative
geometry~\cite{ref-mocioiu,ref-carroll3}, Lorentz violation through
spacetime-varying couplings~\cite{ref-kost20}, and anomalous breaking of
Lorentz and CPT symmetries~\cite{ref-klinkhamer}.

To date, experimental tests of Lorentz violation
have included studies of matter-antimatter asymmetries for
trapped charged particles~\cite{ref-bluhm1,ref-bluhm2,ref-gabirelse,
ref-dehmelt1} and bound state systems~\cite{ref-bluhm3,ref-phillips},
determinations of muon properties~\cite{ref-kost8,ref-hughes}, analyses of
the behavior of spin-polarized matter~\cite{ref-kost9,ref-heckel},
frequency standard comparisons~\cite{ref-berglund,ref-kost6,ref-bear,ref-wolf},
Michelson-Morley experiments with cryogenic resonators~\cite{ref-muller,
ref-stanwix}, Doppler effect measurements~\cite{ref-saathoff,ref-lane1},
measurements of neutral meson
oscillations~\cite{ref-kost10,ref-kost7,ref-hsiung,ref-abe},
polarization measurements on the light from distant galaxies~\cite{ref-carroll1,
ref-carroll2,ref-kost11}, and others. In order to evaluate the results of these
experiments, it has been useful to develop a local quantum field theory that
parameterizes the possible Lorentz violations.
The most general such theory is the standard model extension
(SME)~\cite{ref-kost1,ref-kost2,ref-kost12}. The structure, as well as
stability~\cite{ref-kost3} and
renormalizability~\cite{ref-kost4}, of this extension have been
extensively studied. Most SME calculations are only done to first order in the
Lorentz-violating coefficients, because practically, all Lorentz violations are
small. Few results that are valid to all orders in the Lorentz violation are
known, and new all-orders results, such as those we shall obtain here, are
theoretically interesting.

Another area where high-precision experimental tests are important is in searching
for new forces. Many of the same theories that
predict Lorentz violation also predict that new forces may arise at, for example,
submillimeter distances.
Searches for weak, macroscopic forces other than electromagnetism
and gravity have thus far not been successful. (For recent results,
see~\cite{ref-long,ref-chiaverini,ref-hoyle}. There is, however,
a hint of possible new forces acting
in the Pioneer anomaly~\cite{ref-anderson1,ref-anderson2}.)
Possible new interactions are frequently parameterized
in terms of a Yukawa-type potential. While this kind of potential
is sometimes chosen merely for its simplicity, it does have
strong theoretical motivations,
and it would be worthwhile to consider just what kinds
of Yukawa-like interactions can exist within generalizations of the standard
model. There are also Yukawa interactions in the standard model itself, both
effective (as between pions and nucleons) and fundamental (involving the Higgs
field). So it is reasonable to ask how Lorentz violation might affect
Yukawa physics.

We shall examine this question using the effective field theory approach of the
SME. First, we develop the nonrelativistic Yukawa theory, as modified by
Lorentz violation. We pay particular attention to the potential energy, because
this is what is generally constrained by precision macroscopic force experiments.
Then we
compare the Lorentz-violating Yukawa potential to the analogous quantity that
arises when there is Lorentz violation in the electromagnetic sector.

We shall consider a theory with Lorentz violation in the bosonic sector only.
The Lagrange density is
\begin{equation}
\label{eq-L}
{\cal L}=\bar{\psi}_{A}(i\!\!\not\!\partial-M)\psi_{A}+\frac{1}{2}(\partial^{\mu}
\phi)(\partial_{\mu}\phi)+\frac{1}{2}k^{\nu\mu}(\partial_{\nu}\phi)
(\partial_{\mu}\phi)-\frac{1}{2}m^{2}\phi^{2}-g\bar{\psi}_{A}\psi_{A}\phi.
\end{equation}
A $\phi^{4}$ interaction would also be required for renormalizability; however,
we shall neglect it, as we shall only be considering tree-level phenomena.
The index on the fermion field
$\psi_{A}$ labels distinguishable fermion species. These are included
so that we may extract the interparticle potentials simply, without having to
worry about exchange or annihilation scattering.

The coefficient $k^{\nu\mu}$ is the source of the Lorentz violation. $k^{\nu\mu}$ is
a traceless
symmetric tensor, which modifies the propagation of free scalar particles. This
is the only superficially renormalizable form of Lorentz violation that can be
written down involving only the real scalar field $\phi$.
If the elements of
$k^{\nu\mu}$ have typical size $m/M_{P}$, where $M_{P}$ is some large
scale, then this theory is usable up to scales of order $\sqrt{mM_{P}}$. Above
that scale, one encounters causality violations unless higher-derivative
operators are included in the theory.

For a complex scalar
field $\Phi$, there is also a CPT-odd Lorentz-violating interaction, which
has Lagrange density
\begin{equation}
{\cal L}_{a}=\left(\partial^{\mu}\Phi^{*}\right)
\left(\partial_{\mu}\Phi\right)+ia^{\mu}\left[\Phi^{*}\left(
\partial_{\mu}\Phi\right)-\left(\partial_{\mu}\Phi^{*}\right)\Phi\right]-m^{2}
\Phi^{*}\Phi.
\end{equation}
However, the current term
multiplying $a$ degenerates to zero in the real case, and it is not
natural to consider a complex scalar
field in connection with a Yukawa-type interaction,
because of the theory's Hermiticity
and symmetry structure. A theory with a multiplet of
$2N$ real scalar fields can also be written in terms of $N$ complex scalar fields,
but an $a$-like term will violate the $SO(2N)$ symmetry.

In (\ref{eq-L}), we are neglecting Lorentz
violations in the fermion sector, which are generally much more complicated. Because
of the fermions' Dirac matrix structure,
there are many more Lorentz-violating interactions possible.  However, it
is important to keep in mind that we are treating the most general possible
bosonic Lorentz violation in this theory. Since any new
long-range Yukawa interactions are
known to be quite weak, the Lorentz violation associated with them does not have
to be extremely strongly suppressed to have escaped detection. On the other hand,
Lorentz violations for the most commonly observed fermions---electrons, muons,
protons, and neutrons---have been relatively well constrained.

Moreover, the simple fact that we shall
want to calculate a potential for this theory is actually another
reason to consider
Lorentz violation in the Yukawa boson sector only. Without rotation and Galilean
invariance, the conventional definition of the
nonrelativistic potential may not be useful, as
Lorentz violations can affect the division of the
effective Hamiltonian into potential and
kinetic components in a nontrivial fashion. This fact can be made evident most
easily by looking at the relativistic bosonic system described by ${\cal L}_{a}$.
The action in this theory is a bilinear function of $\Phi^{*}$ and $\Phi$, and
so the theory is free. In particular, a field redefinition
\begin{equation}
\label{eq-atrans}
\Phi\rightarrow e^{ia\cdot x}\Phi,\,\Phi^{*}\rightarrow e^{-ia\cdot x}\Phi^{*}
\end{equation}
converts the Lagrange density into
\begin{equation}
{\cal L}_{a}'=\left(\partial^{\mu}\Phi^{*}\right)\left(\partial_{\mu}\Phi\right)
-(m^{2}+a^{2})\Phi^{*}\Phi.
\end{equation}
So even though $a$ couples to an expression containing derivatives, it can be
seen as affecting the potential part of the Lagrangian. Normally, $a^{2}$ would
be considered a small correction to the mass term, but
for large values of $a^{2}$, the nature of the theory can change qualitatively.
There exists a physical spectrum only for $a^{2}\geq-m^{2}$, for
if $a^{2}<-m^{2}$, the energy is not bounded below. While this bosonic example
is the simplest, similar considerations apply to fermions and in the
nonrelativistic regime. When there is Lorentz violation,
particle energies possess unorthodox structures, and so
their division into kinetic and potential
parts can be tricky and is not necessarily unambiguous. However,
these problems with defining an interparticle potential do not exist if there is
no Lorentz violation in the Lagrangian for the external particles in which we are
interested. (The complexities of the energy-momentum relations in Lorentz-violating
fermion theories are discussed further
in~\cite{ref-kost1,ref-altschul2,ref-altschul3}, where particular attention is
paid to the intimately related question of how the velocity behaves in these
theories.)

The greatest virtue of the simple theory defined by (\ref{eq-L})
is that a number of tree-level results may be worked
out exactly (i.e.\ to all orders in $k^{\nu\mu}$). The Yukawa potential will
illustrate this. This potential will naturally  be
modified by the Lorentz violation, and we shall find that it is a relatively
simple matter to determine the structure of these modifications nonperturbatively
in $k^{\nu\mu}$.

To determine the potential, we consider the nonrelativistic scattering of two
distinguishable
fermions by one-$\phi$ exchange. The particles have initial (final) momenta
$p$ and $k$ ($p'$ and $k'$), and nonrelativistically these must take the form
\begin{equation}
\begin{array}{cc}
p=\left(M,\vec{p}\,\right), & k =\left(M,\vec{k}\right), \\
p'=\left(M,\vec{p}\,'\right), & k' =\left(M,\vec{k}'\right).
\end{array}
\end{equation}
So the momentum exchanged is
\begin{equation}
p'-p=q=\left(0,\vec{p}\,'-\vec{p}\,\right),
\end{equation}
up to corrections suppressed by powers of $|\vec{p}\,|/M$.
Since we are considering a nonrelativistic system, only the $k_{jl}$ portion of
$k^{\nu\mu}$ (which breaks rotational invariance) will contribute.
Any effects of $k_{0j}$ are also suppressed by factors of $|\vec{p}\,|/M$.

The matrix element for the scattering process is then
\begin{equation}
i{\cal M}=\frac{-ig^{2}}{q^{\mu}q_{\mu}+k^{\nu\mu}q_{\nu}q_{\mu}-m^{2}}
\left(2M\delta^{ss'}\right)\left(2M\delta^{rr'}\right),
\end{equation}
where $s$ and $r$ ($s'$ and $r'$) label the incoming (outgoing) spins. The
key quantity in this expression is the part that comes from the Yukawa boson
propagator. This is essentially the three-dimensional Fourier transform of the
modified potential:
\begin{equation}
\tilde{V}\!\left(\vec{q}\,\right)=\frac{-g^{2}}{q_{j}q_{j}-k_{jl}q_{j}q_{l}+m^{2}}.
\end{equation}

To get the interparticle potential, we simply invert the Fourier transform,
\begin{equation}
V\!\left(\vec{r}\,\right)=\int\frac{d^{3}q}{(2\pi)^3}\frac{-g^{2}}
{q_{j}q_{j}-k_{jl}q_{j}q_{l}+m^{2}}e^{iq_{j}r_{j}}.
\end{equation}
We can reduce this integral to the one that appears in the Lorentz-invariant
case by making changes of variables. First, let us define a matrix $K$ with
elements $K_{jl}=\delta_{jl}-
k_{jl}$. This matrix is symmetric and, presuming $|k_{jl}|\ll 1$, positive
definite, so by making an orthogonal rotation of the coordinates, we may reduce it
to a diagonal matrix. Then in the rotated coordinates, the denominator in the
Fourier transform expression becomes
$K_{11}q_{1}^{2}+K_{22}q_{2}^{2}+K_{33}q_{3}^{2}+m^{2}$. We may then rescale the
integration variables by introducing $\bar{q}_{j}=\sqrt{K_{jj}}q_{j}$. (Here and
for the rest of this paragraph, there will be no implied summations over the
index $j$; other repeated Roman indices are still summed, however.)
The exponent becomes $e^{i\bar{q}_{l}\bar{r}_{l}}$, with $\bar{r}_{j}=r_{j}/
\sqrt{K_{jj}}$, and the integration measure transforms as
\begin{equation}
d^{3}q=\frac{d^{3}\bar{q}}{\sqrt{\det K}}.
\end{equation}
This brings $V\!\left(\vec{r}\,\right)$ into the form
\begin{equation}
V\!\left(\vec{r}\,\right)=\frac{1}{\sqrt{\det K}}
\int\frac{d^{3}\bar{q}}{(2\pi)^3}\frac{-g^{2}}{\bar{q}_{l}\bar{q}_{l}+m^{2}}
e^{i\bar{q}_{l}\bar{r}_{l}},
\end{equation}
and the calculation is then elementary:
\begin{equation}
\label{eq-Vrbar}
V\!\left(\vec{r}\,\right)=-\frac{g^{2}}{4\pi\sqrt{\det K}}
\frac{e^{-m\bar{r}}}{\bar{r}},
\end{equation}
where
\begin{eqnarray}
\bar{r} & = & \sqrt{\bar{r}_{l}\bar{r}_{l}} \\
\label{eq-rbar}
& = & \sqrt{\left(K^{-1}\right)_{ln}r_{l}r_{n}}.
\end{eqnarray}

Since (\ref{eq-Vrbar}) and (\ref{eq-rbar}) are written in terms of the
basis-independent quantities $\det K$ and
$\left(K^{-1}\right)_{jl}r_{j}r_{l}$ (now summing over $j$ again),
these expressions are valid in any basis.
In particular, they hold in the original basis in which $K_{jl}$ is not
necessarily diagonal.
So they give an expression for the Lorentz-violation-modified Yukawa potential
which is correct to all orders in $k_{jl}$.

As a simple check of the correctness of this result, we may consider the case in
which $k^{\nu\mu}=Cg^{\nu\mu}$. Then there is no physical
Lorentz violation, and $k^{\nu\mu}$ may be eliminated by rescaling the field $\phi$.
The net result is a rescaling of $g$ and $m$ each by a factor of $1/\sqrt{1+C}$,
which is just what (\ref{eq-Vrbar}) and (\ref{eq-rbar})
together indicate in this instance.

Obviously, it is useful to expand the expression for $V\left(\vec{r}\,\right)$
to leading order in $k_{jl}$. The
determinant part is easy to expand. To leading order, $\det K$ is
equal to $1-k_{jj}$, so
\begin{equation}
\frac{1}{\sqrt{\det K}}\approx 1+\frac{1}{2}k_{jj},
\end{equation}
and this
depends only on the trace $k_{jj}$ over the space part of $k^{\nu\mu}$. However,
this dependence on the trace can simply be absorbed into the overall scaling of
$g$.
The expansion of $\bar{r}$ to leading order in $k_{jl}$ is also simple. Since
$\left(K^{-1}\right)_{jl}\approx\delta_{jl}+k_{jl}$, we have
\begin{equation}
\bar{r}\approx r\left(1+\frac{1}{2}k_{jl}\hat{r}_{j}\hat{r}_{l}\right),
\end{equation}
where $\hat{r}$ is the unit vector $\vec{r}/r$ in the direction
of $\vec{r}$. So the complete ${\cal O}(k^{\nu\mu})$ expression for
$V\!\left(\vec{r}\,\right)$ is
\begin{equation}
V\!\left(\vec{r}\,\right)\approx-\frac{g^{2}}{4\pi}\frac{e^{-mr}}{r}
\left[1+\frac{1}{2}
k_{jj}-\frac{1}{2}\left(1+mr\right)k_{jl}\hat{r}_{j}\hat{r}_{l}\right].
\end{equation}
Since $\vec{r}$ is the only vector appearing in the problem that can be contracted
with $k_{jk}$, the expressions $k_{jj}$ and $k_{jl}\hat{r}_{j}\hat{r}_{l}$ are
really the only tensor structures
that can arise at this order. As previously noted, the first
of these quantities only scales the interaction strength (as long as we
consider effects
only in a fixed reference frame); however,
the second quantity enters in a particular and interesting
fashion. At small distances (less than the boson Compton wavelength), the constant
term in $(1+mr)$ dominates, but at larger distances, the $mr$ part gives the
dominant Lorentz-violating contribution unless the scalar field is massless.

The particle-antiparticle potential is the same. This follows from the fact that
the fermionic structure in the theory is
completely conventional. So we have shown that (\ref{eq-Vrbar}) is essentially the
only interparticle potential that can arise from a renormalizable
Lorentz-violating modification of the Yukawa sector. At leading order, rotation
invariance is broken only by a term of the form $k_{jl}\hat{r}_{j}\hat{r}_{l}$,
while purely phenomenalistic modifications of the Yukawa potential could also
involve the structure $v_{j}\hat{r}_{j}$ (for some constant vector $\vec{v}$) or
similar contractions with tensors of more than two indices.

We can also easily derive the nonrelativistic scattering cross-section from
our expression for $i{\cal M}$. In the center of mass frame, with the two
initial particles approaching one-another along the $z$-axis, each with
(nonrelativistic) energy $\epsilon$, the momentum exchange is just
\begin{equation}
\vec{q}=\sqrt{2M\epsilon}\,(\cos\varphi\sin\theta,\sin\varphi\sin\theta,\cos\theta
-1)=2\sqrt{2M\epsilon}\left|\sin\left(\theta/2\right)\right|\hat{q},
\end{equation}
in terms of the scattering angles $\theta$ and $\phi$.
So the cross section for scattering with the particle spins unchanged is
\begin{equation}
\frac{d\sigma}{d\Omega}=\frac{g^{2}}{256\pi^{2}\epsilon^{2}\left[4\sin^{2}\left(
\theta/2\right)\left(1-k_{jl}\hat{q}_{j}\hat{q}_{l}\right)+\frac{m^{2}}{2M\epsilon}
\right]^{2}},
\end{equation}
and to leading order in $k^{\nu\mu}$, this is just
\begin{equation}
\frac{d\sigma}{d\Omega}\approx\frac{g^{2}}{256\pi^{2}\epsilon^{2}\left[4\sin^{2}
\left(\theta/2\right)+\frac{m^{2}}{2M\epsilon}\right]^{2}}\left[1+\frac
{8\sin^{2}\left(\theta/2\right)k_{jl}\hat{q}_{j}\hat{q}_{l}}{4\sin^{2}
\left(\theta/2\right)+\frac{m^{2}}{2M\epsilon}}\right].
\end{equation}

Since for small enough $m$, the Yukawa potential has bound states, we can also see
how these are affected by the Lorentz violation. Our results still apply in the
$m=0$ limit, and in this limit, we can extract
the leading-order energy shifts for the nonrelativistic bound states by ordinary
perturbation theory. However, because of the degeneracies of the hydrogenic
spectrum, there is a secular matrix to diagonalize, and this cannot generally
be done in closed form. The energy shift has a simple form only for the ground
state, where it is equal to
\begin{equation}
E=-\frac{M}{2}\left(\frac{g^{2}}{4\pi}\right)^{2}\left(1+\frac{2}{3}
k_{jj}\right).
\end{equation}

The potential that we have found is similar in structure to
the modified electrostatic potential $A^{0}$
that appears in a CPT-even Lorentz-violating electromagnetic theory. In such a
theory, the Lorentz violation arises from a Lagrange density
\begin{equation}
{\cal L}_{F}=-\frac{1}{4}F^{\mu\nu}F_{\mu\nu}
-\frac{1}{4}k_{F}^{\mu\nu\rho\sigma}F_{\mu\nu}F_{\rho\sigma}
-A^{\mu}j_{\mu}.
\end{equation}
The four-vector potential for a point charge $q$ (appropriately normalized) in this
theory is~\cite{ref-kost17}
\begin{eqnarray}
\label{eq-A0}
A^{0}\!\left(\vec{r}\,\right) & \approx & \frac{q}{4\pi}\frac{1}{r}
\left[1-\left(k_{F}\right)_{0j0l}\hat{r}_{j}\hat{r}_{l}\right], \\
A_{j}\!\left(\vec{r}\,\right) & \approx & \frac{q}{4\pi}\frac{1}{r}
\left[\left(k_{F}\right)_{0ljl}-\left(k_{F}\right)_{jl0n}\hat{r}_{l}\hat{r}_{n}
\right].
\end{eqnarray}
The angular dependences of $A^{0}\!\left(\vec{r}\,\right)$ and
$V\!\left(\vec{r}\,\right)$ are essentially the same, but what distinguishes the
vector theory is the presence of a nonzero
$\vec{A}$ (and correspondingly, a nonzero $\vec{B}$), even in the absence of
moving charges. This mixing of electrostatic and magnetostatic effects gives us
the possibility of distinguishing, via magnetic measurements, between scalar-
and vector-mediated Lorentz-violating forces. The specific Lorentz-violating
coefficient $\left(k_{F}\right)_{0j0l}$ that appears in (\ref{eq-A0}) can actually
be (to leading order) exported to the matter sector by a redefinition of the
coordinates~\cite{ref-kost17}. This results in a proportional $k_{jl}$ coefficient
for each of the matter fields, but this differs from the Yukawa case we have
considered, because the fermions will also have the same
$k_{jl}$ as the scalar bosons.

If, in the future, weak new forces are discovered, then it is definitely an
interesting question whether these forces are Lorentz invariant. We have shown
that a Lorentz-violating, renormalizable effective field theory can support only
a limited class of modifications to the
nonrelativistic Yukawa potential. Lorentz violations in
the bosonic sector can only result in the specific CPT-even changes to the
potential
that are described by (\ref{eq-Vrbar}). This expression
for $V\!\left(\vec{r}\,\right)$ is correct to all orders in
the Lorentz violation, but at leading order it closely resembles the electrostatic
potential in a Lorentz-violating electromagnetic theory. However, the
electromagnetic theory will generally also contain mixing between electric and
magnetic interactions, which distinguishes it from the scalar Yukawa theory.
Identifying the field-theoretical origins of any experimentally observed Lorentz
violations would be very important, and these results provide the tools to help
make such an identification.

\section*{Acknowledgments}
This work is supported in part by funds provided by the U. S.
Department of Energy (D.O.E.) under cooperative research agreement
DE-FG02-91ER40661.

\end{document}